# Seeing Site-Specific Isotopic Labeling of Amino Acids with Vibrational Spectroscopy in the Electron Microscope

*Jordan A. Hachtel[1*], Jingsong Huang[1,2], Ilja Popovs[3], Santa Jansone-Popova[3], Jacek Jakowski[1,2] and Juan Carlos Idrobo[1]*

[1] Center for Nanophase Materials Science, Oak Ridge National Laboratory, Oak Ridge, TN 37831 USA
[2] Computational Sciences and Engineering Division, Oak Ridge National Laboratory, Oak Ridge, TN 37831, USA
[3] Chemical Sciences Division, Oak Ridge National Laboratory, Oak Ridge, TN 37831 USA

## *Abstract*

Isotope labeling is a fundamental staple for the study of cellular metabolism and protein function. The conventional techniques that allow resolution and identification of isotopically-labeled biomarkers, such as mass spectrometry and infrared spectroscopy, are macroscopic in nature and have the disadvantage of requiring relatively large quantities of material and lacking spatial resolution. Here, we record the vibrational spectra of an α-amino acid, L-alanine, using spatially-resolved monochromated electron energy loss spectroscopy (EELS) to directly resolve carbon-site-specific isotopic labels in a scanning transmission electron microscope. The EELS is acquired in aloof mode, meaning the probe is positioned away from the sample (~20 nm) sparing the sensitive biomolecule from the high-energy excitations, while the vibrational modes are investigated. An isotopic red-shift of 5.3 meV was obtained for the C=O stretching mode in the carboxylic acid group for $^{13}$C-enriched L-alanine when compared with naturally occurring $^{12}$C L-alanine, which is confirmed by macroscopic infrared spectroscopy measurements and theoretical calculations. The EELS experiments presented here are the first demonstration of non-destructive resolution and identification of isotopically-labeled amino acids in the electron microscope, opening a new door for the study of biological matter at the nanoscale.

**Keywords:** Vibrational EELS, isotope-labeling, amino acids, biomolecules

*Introduction*

The ability to detect and identify proteins with isotopically-labeled sites is a vastly important research topic in life science, and especially in metabolomics and proteomics (*1–3*). The most frequently used technique for this type of analysis is mass spectrometry, where the mass-to-charge-ratio of ionized molecules can be used to accurately determine the atomic weight and isotopic composition of the fragments. However, the sample is destroyed by the experiment, leading to the loss of valuable information pertaining to higher order structure and associated supramolecular interactions (*4*). Alternatively, specific isotopes can be observed through frequency changes of the molecular vibrational modes corresponding to the difference in atomic weights. Thus, isotopic analysis can be conducted in spectroscopy techniques that can detect these shifts, such as Fourier-transform infrared spectroscopy (FTIR) (*5*), Raman spectroscopy (*6*), and inelastic neutron scattering (*7*). Additionally, vibrational spectroscopy has the advantage that it is generally non-destructive and highly sensitive to the atomic structure, allowing for the use of isotopes as biomarkers for direct visualization and observation of dynamic changes in biomolecules (*8–10*). However, the conventional techniques are generally macroscopic experiments and require large quantities of a sample to be examined statistically. As such, results obtained from these techniques are insensitive to local variations in the vibrational signatures within the biomolecules, presenting an inherent need for techniques possessing simultaneous high spatial and spectral resolution.

Scanning probe optical techniques such as tip-enhanced Raman spectroscopy (TERS) and scanning near-field optical microscopy (SNOM) have demonstrated the capacity to examine the vibrational spectra of biomolecules with high spatial resolution (*11–15*). However, as surface techniques, TERS and SNOM are limited to specific sample geometries and highly sensitive to surface states. Electron microscopy is another spatially-resolved technique that has already shown promising applications in the life sciences. Cryogenic electron microscopy has recently emerged as a powerful method to study biological matter with high spatial resolution, yielding remarkable new insights in structural biology and even garnering the 2017 Nobel Prize in Chemistry (*16–18*).

Isotopic analysis has been conducted in the electron microscope via electron scattering from crystalline solids (*19*), but this technique cannot be directly transposed to amino acids or other biological materials as the high energy electrons in the beam instantly ablate sensitive organic samples. Electron energy-loss spectroscopy (EELS) has historically been used to acquire vibrational spectra, but only in the low-electron-energy reflectance geometry, which is experimentally limited to surfaces and is not spatially-resolved (*20*, *21*). However, recent breakthroughs in electron monochromation have opened the door to vibrational EELS in the scanning transmission electron microscope (STEM), improving energy resolution and reducing background such that phonons in solids can be measured with high spatial resolution (*22–25*). Furthermore, the vibrational spectra can be acquired with the electron probe positioned near to, but not in contact with, the sample. In this 'aloof' acquisition mode, the vibrations are measured via the coupling with the evanescent field of the electrons (*26*). Aloof vibrational EELS can be recorded at room temperature with high efficiency, meaning that biomolecules can now be analyzed in the electron microscope without the need for cryogenic conditions and without damaging the specimen (*27–29*) opening the path of detection of isotope labeling in the electron microscope.

Here, we demonstrate for the first time the detection and determination of site-specific isotope labeling in an amino acid obtained from an electron microscope. We examine L-alanine and its $^{13}$C labeled counterpart using aloof monochromated EELS in an aberration-corrected STEM. An isotopic shift of 5.3 meV is measured with EELS, which is consistent with macroscopic FTIR experiments. Additionally, density functional theory (DFT) calculations are performed to identify the specific vibrational modes shifted between the $^{12}$C and $^{13}$C alanine samples, which are found to primarily originate from the carboxylic acid site. These results demonstrate that spatially-resolved site-specific isotopic vibrational analysis at the nanometer scale can be conducted on biological samples in the electron microscope.

## *Results and Discussion*

The samples are prepared by crushing and dispersing the high purity powders onto TEM grids, leaving small crystalline clusters of L-alanine (sizes varying between hundreds of nm and tens of microns) scattered across a lacey carbon support. The lacey carbon grid can potentially have an extremely weak C-H stretch vibrational mode from adsorbed hydrocarbons (*29*), but the signal originating from the L-alanine dominates the vibrational spectrum due to the significantly higher thickness of the sample.

The aloof vibrational electron energy-loss (EEL) spectrum of L-alanine, across the mid-infrared regime is shown in Figure 1 (blue), featuring peaks ranging from 100-400 meV (~800-3,200 cm$^{-1}$), including a broad wedge-shaped peak from 300-400 meV, two sharply pronounced peaks between 160 and 210 meV, along with several smaller peaks at energies lower than 160 meV and a small bump at 260 meV. To compare the EEL vibrational spectrum to conventional techniques,

FTIR was performed on a sample prepared from the same material and is plotted in Fig. 1 (red). The EELS and FTIR spectra in Figure 1 are acquired with energy resolutions of 13 meV and 1 meV, respectively. As a result, a large number of peaks are observed in FTIR in regions where a fewer number of broader peaks are observed in EELS. To provide a better match to the energy resolution of EELS, a 13 meV full-width at half-maximum (FWHM) Gaussian blur is applied to the FTIR spectra, and plotted next to the as-acquired FTIR and EELS in Fig. 1 (green). In the broadened FTIR spectrum the vibrational peak frequencies and intensities match extremely well with the EELS.

To understand which vibrational modes in alanine produce the infrared response of the molecule, first-principles DFT calculations were carried out for the fundamental vibrations of an L-alanine molecule in its global minimum structure with and without isotope substitutions. The eigenvalues of all the specific vibrational modes are plotted with respect to their intensities in Figure 2. Similarly, a 13-meV-FWHM Gaussian broadening was performed on the eigenvalues to produce a theoretical vibrational spectrum at the same energy resolution as the EELS experiments. Figure 2a shows vibrational eigenvalues and spectrum of the L-alanine molecule with the carbon atoms in their naturally occurring $^{12}$C isotope. The higher energy peaks (~350-450 meV/2,800-3,600 cm$^{-1}$) correspond to the hydrogen stretching modes, with strong contributions from the O-H and C-H modes, but weaker ones from N-H modes. At lower energies, there are two dominant peaks, one at 216 meV/1,745 cm$^{-1}$ that originates from the C=O stretching mode, and another at 136 meV/1,095 cm$^{-1}$ that originates from the stretching of CO-O in the carboxylate group. Below the stretching modes, the O-H and N-H bending modes dominate the lowest energies of the spectrum at 73 meV/590 cm$^{-1}$ and 107 meV/865 cm$^{-1}$ respectively.

A comparison between the theoretical spectrum in Figure 2a with the experimental spectra in Figure 1 indicates that the standalone peaks for the theoretical O-H and C-H stretching modes differ from the experimentally observed wedge-shaped peak from 300-400 meV. The difference should be ascribed to the fact that the OH group in the carboxylic acid chain tends to form intermolecular hydrogen bonds (*30*), which are known to reduce and broaden the vibrational frequencies of carboxylic group in a condensed phase when compared to individual molecules. In addition, the calculations show that the dominant peak in the FTIR and EELS at ~200 meV should come from the C=O stretching mode calculated at 216 meV. DFT is known to overestimate the frequencies of optical excitations, but the offset is found to be small even though the calculation is for one individual molecule while experiments were done for a condensed phase. Further, it is possible that the secondary peak observed in EELS is the CO-O stretching mode, but there is a slightly larger offset (136 meV theoretical vs. 170 meV experimental), and the calculated frequency is also at a lower energy than the experimental peak while most of the other peaks are at higher energies than their experimental counterparts. The difference could also potentially be due to the formation of intermolecular hydrogen bonds, which may increase the force constant of the CO-O bond by a partial double bond character. The fundamental vibrational response is recalculated upon $^{13}$C enrichment in the methyl group (Fig. 2b), the amino group (Fig. 2c), and the carboxylate group (Fig. 2d). For the methyl and amino group sites, changes in the vibrational frequencies can be observed across the spectrum, but the changes are small and no significant changes to the Gaussian-broadened spectra are detected. However, when the carboxylic acid group has a $^{13}$C isotope, the vibrational spectrum exhibits a significant redshift for the main C=O stretch mode that manifests itself clearly even in the broadened spectrum. The redshift by a factor of 1.025 (theory) or 1.029 (experiment) can be rationalized by the change of reduce mass of C=O group by the factor of 1.023 according to $\sqrt{\mu^{13}CO/\mu^{12}CO}$. All other differences in the vibrational spectra are minimal, and almost entirely negligible once broadened, indicating that any spectral changes originate specifically from the carboxylic acid $^{13}$C enrichment.

Figure 3a and 3b shows the EEL spectra from the $^{12}$C (Fig. 3a) and $^{13}$C (Fig. 3b) enriched L-alanine. In order to obtain a precise measurement of the peak position of the C=O stretching mode, a two-Lorentzian fit of the double peak structure between 160 meV and 210 meV is (also plotted in Figure 3a and 3b). It can be seen clearly that a significant shift is observed. In order to obtain a high precision and accuracy measurement, 300 such spectra are acquired and fitted for each sample. The averages and distributions for the measurements and peak fits are shown in Figure 3c. The average energy position of the higher energy C=O peak is measured to be 199.2 meV for the $^{12}$C and at 193.9 meV for $^{13}$C (a shift of 5.3 meV), with standard deviations of 0.8 meV and 1.2 meV, respectively. The results demonstrate that a significant isotopic shift has been measured with EELS with high precision, and that even smaller shifts could still be clearly resolved with vibrational EELS.

To compare the measured isotopic shift in EELS to theoretical calculations, it is important to include the conformational isomers of L-alanine. The electron beam samples a statistical number of molecules in the alanine cluster, meaning that the beam detects all the conformers present and they contribute to the vibrational spectrum. To accommodate the alanine conformers in the first-principles calculations, the vibrational response of each conformer is calculated and weighted by a Boltzmann factor to estimate the fraction of alanine molecules in that specific structure. The conformer-weighted theoretical vibrational spectrum for both $^{12}$C and $^{13}$C is shown in Figure 3d.

The theoretical predicted isotopic shift is 5.4 meV, agreeing extremely well with the 5.3 meV EELS measurements.

The isotopic shift is also measurable via FTIR, shown for the as-acquired spectrum in Figure 3e, and the Gaussian-broadened spectrum in Figure 3f. While isotopic shift in the as-acquired spectra is 5.0 meV, the broadening increases the measured shift to 5.7 meV due to the fine structure of the dominant C=O peak. The isotopic shifts measured for the two FTIR resolutions are centered on the EELS value of 5.3 meV indicating an excellent experimental match between FTIR and EELS. The averaged EEL spectrum is shown in Figure 3g across the same energy range as the plots in Figures 3d-3f.

*Conclusions*

The ability to measure isotopic shifts of site-specific vibrational modes in amino acids opens up a wide range of possibilities for nanoscale biological experiments via monochromated EELS in the STEM. The achievement of isotopic sensitivity for amino acids in the electron microscope is an exciting step forward as the conventional electron methodologies (such as EELS, X-ray dispersive spectroscopy, and high-angle electron scattering) are either insensitive to isotopic changes or incompatible with organic materials. Additionally, the capacity to produce spatially-resolved, high-signal-to-noise-ratio, high-energy resolution spectra from minute quantities of organic material makes it a strong complement to other standard techniques. In future experiments, it should be possible to measure isotopic concentrations, creating the possibility to conduct experiments such as nanoscale carbon dating, bringing electron spectroscopy into the forefront of the life sciences.

*Methods*

*First Principles Calculations:* Structural and vibrational calculations for L-alanine were performed at the DFT level with B3LYP/aug-cc-pVDZ using NWChem (*31*). First, twelve conformational isomers, including all of those reported in the literature (*32*), were located by systematically varying the orientations of -NH$_2$, -COOH, and -OH groups, followed by full geometry optimizations. Afterwards, vibrational calculations were performed to verify their identities as local or global minima instead of saddle points. Vibrational calculations also produce the fundamental vibrational frequencies and intensities at the room temperature for all of the conformers. Vibrational frequencies were scaled by a linear factor of 0.97 (*33*), while intensities for the local minima were scaled by their energy differences with the global minimum according to the Boltzmann distribution at room temperature. For isotope effect, the atomic mass of the carbon atoms was set to 12 exactly and 13.003355 for the $^{12}$C and $^{13}$C isotopes, respectively. Finally, by normal mode analysis of the theoretical results, vibrational peaks of interest were assigned in the experimental FTIR spectrum. L-alanine is known to exist in zwitterionic form under biological conditions. Therefore, the conformational space of the zwitterionic form was also sampled. However, it was found that in the single molecular calculations, all of the zwitterionic isomers changed their structures to the amino acid form during geometry optimizations.

*Electron Energy Loss Spectroscopy:* All EEL spectra are acquired at Oak Ridge National Laboratory on a Nion aberration-corrected high energy resolution monochromated EELS-STEM (HERMES$^{TM}$) operated at 60 kV accelerating voltage (*34*). The microscope is equipped with a prototype Nion spectrometer possessing a Hamamatsu ORCA high-speed CCD detector. The energy resolution of the EELS acquisition is taken to be the FWHM of the zero-loss peak (ZLP), or the peak in the EEL spectrum that contains all the counts from electrons that have only elastically scattered from the sample and have lost no energy. The cold field emission gun without monochromation possesses an energy resolution is ~270 meV, but by monochromating the beam the energy resolution can be brought down to ~10 meV allowing for phonons to be resolved directly (*35*). All EEL spectra are acquired with a 1 mm aperture corresponding to a collection angle of 13 mrad, a probe with a convergence semiangle of 30 mrad, and a beam current of ~300 pA. For the aloof acquisitions the spectra were acquired with an impact factor of approximately 20 nm, it was found that for smaller impact parameters even the weak tails of the electron probe damaged the alanine clusters. The measured clusters were ~2 μm in diameter. Additionally, all EEL spectra were produced by acquiring multiple acquisitions at short dwell times, and subsequently using sub-pixel alignment to create a single summed spectrum. The short dwell times are required to minimize the effect of tip-noise on the EELS, to keep the ZLP from saturating the detector and to avoid losing the calibration zero energy point of the ZLP peak maximum. To fit and subtract the backgrounds for isotopic shift measurement in Figure 3, a power law, $I(\Delta E) = I_0 \cdot \Delta E^{-r}$, is used to fit the background between 70 and 90 meV before the strongest peaks in the alanine vibrational spectrum. In order to fit the C=O stretching it is found that a two-Lorentzian fit provides a better match to the data, as the tail of the lower-energy weaker peak influences the higher-energy C=O peak, thus by fitting it with a Lorentzian the true shape of the C=O peak can be fit more clearly. Processing of the EELS was accomplished either with Nion Swift software and using in-house add-on Python scripts.

*FTIR measurements*: As received, commercial L-alanine $^{12}$C and $^{13}$C-enriched samples (Sigma Aldrich) were used as powders for spectra acquisition. The spectra were recorded using PerkinElmer Frontier Fourier transform instrument.

### *Acknowledgments*

This research was supported by the Center for Nanophase Materials Sciences, which is a Department of Energy Office of Science User Facility (J.A.H., J.H., J.J. and J.C.I.), and by the Department of Energy, Office of Science, Basic Energy Sciences, Chemical Sciences, Geosciences, and Biosciences Division (I.P., S.J.P.). This research was conducted, in part, using instrumentation within ORNL's Materials Characterization Core provided by UT-Battelle, LLC under Contract No. DE-AC05-00OR22725 with the U.S. Department of Energy. Theoretical calculations used resources of the National Energy Research Scientific Computing Center, which is supported by the Office of Science of the U.S. Department of Energy under Contract No. DE-AC02-05CH11231.

*Contributions*

J.C.I. conceived the experiment with input from I.P. and S.J.P. I.P. performed the FTIR measurements. J.H. and J.J. performed the theoretical calculations and advised in the experimental data interpretation. J.A.H. performed the experiments, analyzed the data, prepared the figures, and wrote the main text of the manuscript. J.C.I. advised during the preparation of the manuscript. All the authors discussed the results and contributed to the revision of the manuscript.

*Competing Interests*

The authors declare no competing interests.

*Corresponding author*

Correspondence to hachtelj@ornl.gov

*Figures*

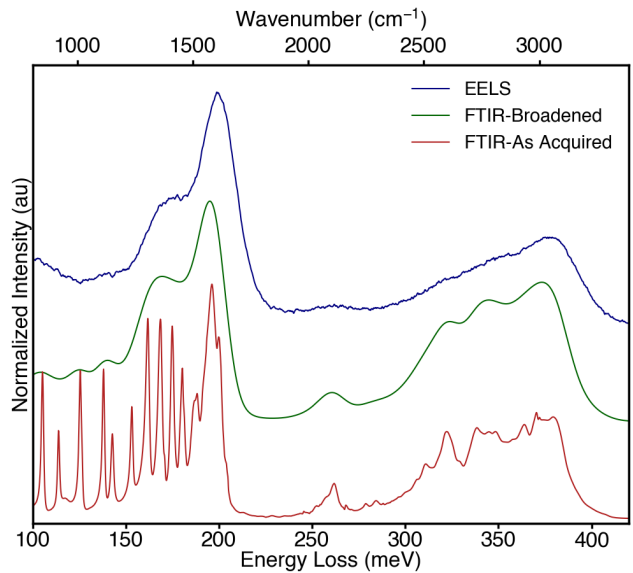

**Figure 1: Vibrational Spectroscopy of L-Alanine.** Experimental vibrational spectra of L-alanine acquired with monochromated 'aloof' EELS (blue), Gaussian-broadened FTIR (green), and FTIR in its as-acquired energy resolution (red). The Gaussian broadening is used to show the FTIR at the same energy resolution as the EELS.

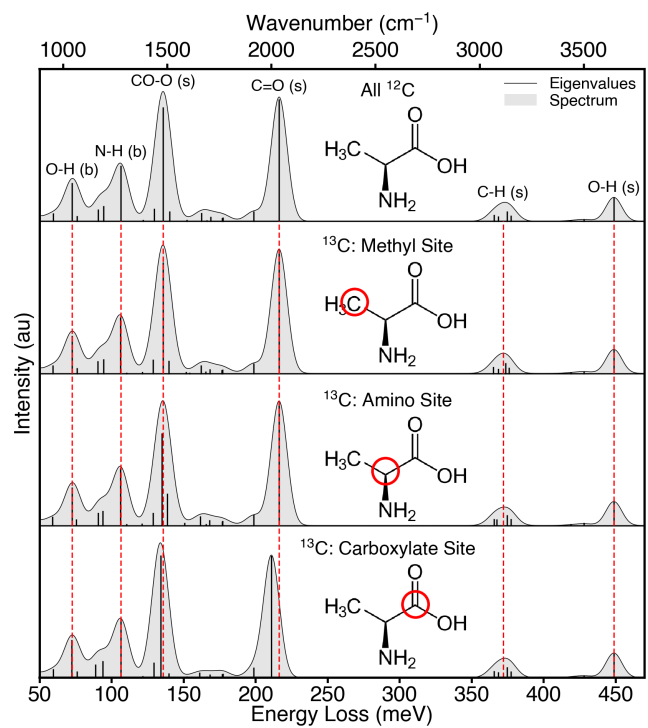

**Figure 2: Influence of Isotopic Enrichment at Carbon Sites.** DFT calculated vibrational eigenvalues and the corresponding Gaussian broadened spectra for the (a) L-alanine (a) and its $^{13}$C isotopic enriched counterparts at each of the three carbon sites: (b) methyl site, (c) amino site, (d) carboxylate site (shown schematically in the insets by red circles). Small shifts are visible in individual eigenvalues but only $^{13}$C at the carboxylate site generates any significant change in the spectra.

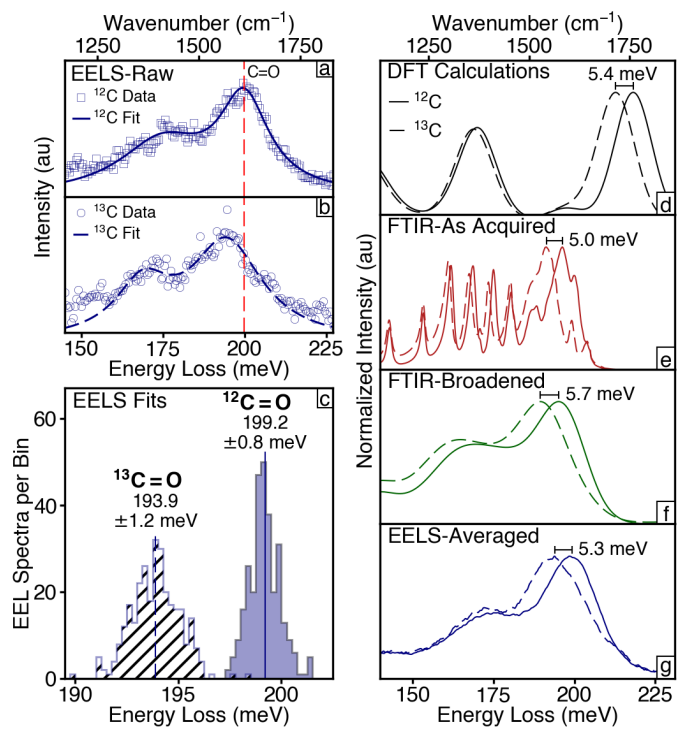

**Figure 3: Measurement of Isotopic Red Shift in Vibrational EELS.** (a) Raw EEL spectrum acquired from (a) $^{12}$C and (b) $^{13}$C L-Alanine samples, with two-Lorentzian fit of spectral region between 160 meV and 210 meV after power-law background subtraction. To perform high precision measurements 300 spectra are acquired and fitted. (c) Histogram of fitted peak positions from all acquisitions in both samples. Measured isotopic peak shifts in DFT (d), FTIR (e), FTIR broadened to EELS energy resolution (f), and in average of 300 EEL spectra (g), demonstrating a high accuracy shift measurement in EELS.